\documentclass[aps,prl,twocolumn,showpacs]{revtex4}
\usepackage{epsfig,here,graphicx}
\usepackage{amsmath}

\newcommand{\rem}[1]{}

\newcommand{\ui}{\mathrm{i}}
\newcommand{\ue}{\mathrm{e}}

\begin{document}

\title{Directional Emission from an Optical Microdisk Resonator with a Point Scatterer}

\author{C. P. Dettmann, G. V. Morozov, M. Sieber, H. Waalkens}
\affiliation{Department of Mathematics, University of Bristol, Bristol BS8 1TW,
United Kingdom}

\date{\today}

\begin{abstract}
We present a new design of dielectric microcavities supporting modes
with large quality factors and highly directional light emission.  The
key idea is to place a point scatterer inside a dielectric circular
microdisk.  We show that, depending on the position and strength of the
scatterer, this leads to strongly directional modes in various frequency
regions while preserving the high Q-factors reminiscent of the
whispering gallery modes of the microdisk without scatterer.  The design
is very appealing due to its simplicity, promising a cleaner
experimental realisation than previously studied microcavity designs on
the one hand and analytic tractability based on Green's function
techniques and self-adjoint extension theory on the other.

\end{abstract}


\maketitle


{\em Introduction.---} Modern fabrication techniques allow one to 
build dielectric optical resonators on a microscopic scale.
Light is trapped by utilizing the  principle of total internal reflection.
These microcavities have great potential
for a wide range of applications and studies in laser physics
and microphotonics \cite{Vahala2003,Ilchenko2006}, like 
the realization of miniature laser sources, the
creation of dynamical filters for optical communications and the
suppression and enhancement of spontaneous emission. 

For many practical purposes it is crucial to have microcavities 
%
%
that possess resonances with long lifetimes (which is a prerequisite for low
threshold lasing) and highly directional emission patterns. 
The lifetimes are characterized by the so called $Q$-factor given by
$Q=\omega/\Delta \omega$, where $\omega -\ui \Delta \omega/2$ is the
complex frequency of the resonance with $\Delta \omega$ being the linewidth or
inverse lifetime.
The best known example of modes with high $Q$-factors are the so called
whispering gallery modes (WGMs) which are the optical analogues of the acoustic waves
evolving along the walls of convex shaped halls first studied by Lord
Rayleigh in the 19th century.
The experimental realization of a 
thin microdisk laser based on WGMs was
first reported in \cite{Logan1992}.
Theoretical studies \cite{Zozoulenko2004} show that
the WGMs of an ideal 
circular microcavity can lead to very  high $Q$-factors of the order $10^6-10^{13}$.
Despite a significant degradation due to imperfections
on the disk boundary, inhomogeneity of the refractive index inside the disk,
effects of coupling to the substrate etc, typical experimental $Q$-factors 
of the dielectric disk resonances remain quite high, usually 
$\approx 10^4$, with the record value of $Q \approx 5 \times 10^5$
\cite{Borselli2004}.
However, the applicability of circular microdisk lasing cavities 
is limited by their isotropic light emission.
In order to obtain a directional optical output 
one has to break the  rotational symmetry, for example, by deforming 
the boundary of the cavity \cite{Logan1993,Stone1997,Gmachl1998}.
This significantly improves the emission directionality but typically
spoils the  $Q$-factors.  
Another approach to breaking the symmetry is to insert an obstacle 
like a linear defect \cite{ApalkovRaikh2004} or a hole~\cite{Wiersig2006} into the
microdisk.
This indeed  allows one to obtain resonances 
with very large $Q$-factors and relatively high directionality. 
However, all the systems mentioned above require quite extensive
numerics like the boundary element method \cite{Wiersig2003} or the $S$-matrix approach
\cite{Hentschel2002,Zozoulenko2004} to find resonances and
optimize design parameters.

\begin{figure}[h]
\centerline{
\includegraphics[width=3cm,height=3cm]{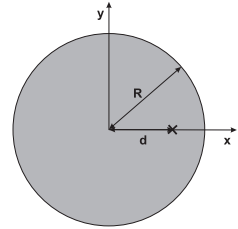}
}
\caption{Microcavity of refractive index $n$ and radius $R$  
with a point scatterer at the distance $d$ from the center. The
external medium has refractive index $n_{\rm {ext}}=1$.} 
\label{Fig1}
\end{figure}

In this Letter, we propose a much simpler method which 
significantly improves directionality 
of the modes of conventional microdisk resonators 
while keeping their $Q$-factors high ($>10^4$).
The symmetry is broken by placing a point scatterer 
within the inner region of the microdisk, see Fig.~\ref{Fig1}.
It turns out that such a geometry improves the emission directionality
of microdisk modes for a wide range of frequencies, 
especially in the visible spectrum.
Moreover, this approach is to a large extent analytically tractable 
enabling a systematic optimization of the design parameters 
(location and strength of the scatterer) 
with only modest numerical effort. 
Closed systems with a point scatterer have been extensively studied in the
context of `quantum chaos' \cite{Seba1990}  showing that the spectral properties of such systems can be
obtained from  self-adjoint extension theory~\cite{Zorbas1980}.
For an open microdisk  we will mainly refer to these results. 
Similarly to the case of closed systems, it turns out that the
resonance wavefunctions of the open microdisk with a point scatterer are
essentially given by the Green's function of the microdisk without scatterer 
(unperturbed microdisk).

{\em Unperturbed Microdisks.---} If we treat a microcavity as a passive object,
we can find its resonances from Maxwell's equations 
with a refractive index independent of the EM field.
For a zero axial momentum EM field, i.e. for  waves with $k_z=0$, 
where $z$ is perpendicular to the disk plane $(x\,y)$,
a thin passive microdisk can be modeled
as a 2D dielectric disk of radius $R$,  
with the effective refractive index $n_{{\rm eff}}(r)=n$,
which takes into account the material as well as the thickness of the microdisk.
In this model, Maxwell's equations reduce 
to two scalar Helmholtz equations corresponding to
TM and TE polarizations, respectively.

In this Letter we consider  only TM modes.
The electric field is then of the form ${\bf E}=E_z(x,y)\,{\bf e_z}$, and for 
a wavenumber $k=\omega/c$, 
$E_z$ satisfies the Helmholtz equation
$(\nabla^2+k^2 n^2) E_z  = 0$.
Due to circular symmetry this equation can be separated in 
polar coordinates $(r,\phi)$ 
where the TM modes are then characterized by  azimuthal
and radial quantum numbers
$m=0,\pm1,\pm2,\ldots$ 
and  $q=1,2,3,\ldots$, respectively.
The corresponding Green's function  
is given by (see, e.g., \cite{Morse1953})
\begin{equation} 
G\left( {{\bf r},{\bf r}_{\bf 0}, k } \right) 
=\sum_{m=-\infty}^\infty \frac{\ue^{i m ({\varphi  - \varphi_{0} } )}}{ 2\pi r_0 W}   E_1(r_{<},k)E_2(r_{>},k)\,, \nonumber
\end{equation}
where  $E_{1,2}(r)$ are solutions of the homogeneous equation
\begin{equation}
\left( \displaystyle\frac{{d^2 }}{{dr^2 }} + \displaystyle\frac{1}{r}\displaystyle\frac{{d}}{{dr}} 
+ {k^2 n^2 (r) - \displaystyle\frac{{m^2 }}{{r^2 }}} \right)E\left( {r} \right) = 0\,, \nonumber
\end{equation}
$W=W(E_1,E_2)$ is the Wronskian evaluated at $r_0$,
and $r_{<}$ ($r_{>}$) is the smaller (larger) of $r$ and $r_0$. 
The physical boundary conditions require $E_1(r)$ to be finite 
at $r=0$ and $E_2(r)$ to be an outgoing wave for $r \rightarrow \infty$.
Requiring $E_{1,2}$ to be smooth at $r=R$ then leads to

\begin{widetext}
\begin{equation} \label{eq:G_is}
G\left( {{\bf r},{\bf r}_0, k} \right) = \left\{ {\begin{array}{*{20}c}
   { - \displaystyle\frac{\ui}{4}H_0 \left( {kn \left| {{\bf r} - {\bf r}_0 }
   \right|} \right) - \displaystyle\frac{\ui}{4}\displaystyle\sum\limits_{m  =
   0}^\infty  {\displaystyle\frac{{C_{m} }}{{A_{m} }} \epsilon_m \cos\left[ m\left( {\varphi  - \varphi _0 } \right)\right]J_m \left( {kn r_<  } \right)J_m  
 \left( {kn r_ >  } \right)} ,} & {r_ <  ,r_ >  \, < \,R,}  \\
   { - \displaystyle\frac{1}{2\pi k R}\displaystyle\sum\limits_{m = 0}^\infty  {\displaystyle\frac{1}{{A_{m} }}\epsilon_m\cos \left[m\left( {\varphi  - \varphi _0 } \right)\right]J_m \left( {kn r_ <  } \right)H_m \left( {k r_ >  } \right)} ,} & {r_ <   < R,\,\,r_>   > R,}  \\
\end{array}} \right.
\end{equation}
\end{widetext}
where $\epsilon_m=2$ if $m\ne0$ and $\epsilon_m=1$ if $m=0$, and
\begin{equation} \nonumber
\begin{split}
A_{m} & =  n H_m \left( {kR} \right)J_m^{\,\,\,'} \left( {knR} \right) - H_m^{\,\,\,'} \left( {k R} \right)J_m \left( {kn R} \right),  \\ 
C_{m}& =  H_m \left( {kn R} \right) H_m^{\,\,\,'} \left( {k R} \right) - n
H_m^{\,\,\,'} \left( {kn R} \right)H_m \left( {kR} \right). 
\end{split}
\end{equation}
The functions $J_m$ and $H_m$ are Bessel and Hankel functions of the first
kind, respectively.
The resonances $k_{\,\rm {res}}$ of the microdisk are given by the
poles of the Green's function (\ref{eq:G_is}), 
i.e. they satisfy the condition
\begin{equation} \label{eq:unpert_quant_cond}
nH_m ( k_{\,\rm {res}} R )J_m^{\,\,\,'} ( k_{\,\rm {res}} nR ) -
H_m^{\,\,\,'} (k_{\,\rm {res}} R) J_m ( k_{\,\rm {res}} n R ) 
= 0.\nonumber
\end{equation}
Resonances differing by the sign of  $m$ are degenerate.


{\em Microdisks with a point scatterer.---}
The Green's function (\ref{eq:G_is}) is logarithmically divergent at the point
${\bf r}={\bf r_0}={\bf d}$ where $d<R$. Since
\begin{equation}
H_0(z) = 1 + \ui \frac{{2}}{\pi }\left( {\ln z + \gamma  - \ln 2} \right) +
{\cal O} (z^2)  \nonumber
\end{equation}
where $\gamma = 0.5772156649\ldots$ is the Euler-Mascheroni constant,
the regularized Green's function $G_{{\rm r}}$ can be  obtained from (\ref{eq:G_is}) 
if we subtract the term ${{\ln \left( {k_0 \left| {{\bf r} - {\bf r}_0 } \right|} \right)} \mathord{\left/
{\vphantom {{\ln \left( {k_0 \left| {{\bf r} - {\bf r}_0 } \right|} \right)} {2\pi }}} \right.
\kern-\nulldelimiterspace} {2\pi }}$, where $k_0$ is an arbitrary constant.
It then follows from self-adjoint extension theory~\cite{Zorbas1980} that
the resonances $k_{\rm {res}}$ of the microdisk with a point scatterer at
position ${\bf d}$ and 
of coupling strength $\lambda$ 
satisfy the condition \cite{Shigehara1994}
\begin{equation}\label{eq:perturbed_quant_cond}
0 = 1 - \lambda \,G_{{\rm r}} ({\bf d},{\bf d}, k_{\,\rm {res}} )\,,
\end{equation}
where $G_{{\rm r}}({\bf d},{\bf d},k)$ is the regularized Green's function at the point
${\bf r}={\bf r}_0={\bf d}$. 
It is convenient to introduce the new coupling parameter $a$ which is defined
by 
${{2\pi } \mathord{\left/{\vphantom {{2\pi } {\lambda  \equiv  - \ln \left( {k_0 a} \right)}}} \right.
\kern-\nulldelimiterspace} {\lambda  \equiv  - \ln \left( {k_0 a} \right)}}$
and hence is a combination of the free parameter $k_0$ and the coupling
strength $\lambda$. Equation~(\ref{eq:perturbed_quant_cond}) then reads
\begin{equation}
0 = -\frac{\ui}{4}+\frac{{1}}{2\pi }\big( {\ln \frac{{k_{\,\rm {res}} n a }}{{2 }} + \gamma } \big)
- \frac{\ui}{4}\sum_{m=0}^\infty {\frac{{C_m }}{{A_m }}} \epsilon_m {J^2_m \left( {k_{\,\rm {res}}nd } \right)}\,.\nonumber
\end{equation}
The parameter $a$ can be directly interpreted in the limit $ka\ll 1$ as the
radius of a localized perturbation of the disk \cite{ExnerSeba96}. 

The resonance wavefunctions  $E_z({\bf r},k_{{\rm res}})$ are  given by 
\begin{equation}
E_z\left( {\bf r},k_{{\rm res}} \right) = N\,G\left( {{\bf r},{\bf d} ,k_{\rm {res}}} \right),\nonumber
\end{equation}
where $G\left( {{\bf r},{\bf d} ,k_{\rm {res}}} \right)$ 
is the unregularized Green's function~(\ref{eq:G_is})
and $N$ is a normalization factor. 

Note that $m$ is no longer a good quantum
number of the microdisk with a point scatterer.
However, it follows from the condition~(\ref{eq:perturbed_quant_cond})
that for a vanishing strength of the point scatterer ($\lambda=0\pm$ or
equivalently $a=0$ or $a=+\infty$) the resonances of the microdisk 
with scatterer coincide with those of the unperturbed microdisk,
which, for $m\ne0$, are twofold degenerate.
In fact, if the coupling parameter $a$ varies from  0 to $+\infty$ 
one of the members of each pair of degenerate resonances remains unchanged.
The other member is moving in the complex $k$ plane along a line segment that
connects two resonances of the unperturbed microdisk.
This can be understood from the fact that for a vanishing strength
of the scatterer the angular dependence of the resonant modes 
is given by $\sin(m\varphi)$ and $\cos(m\varphi)$. 
The unperturbed resonance is the one with a nodal
line along the $x$-axis on which we place the scatterer (see
Fig.~\ref{Fig1}), i.e. the resonance with the angular part  $\sin(m\varphi)$.

\begin{figure}[htp]
\centerline{
\includegraphics[width=8.6cm,height=7.2cm]{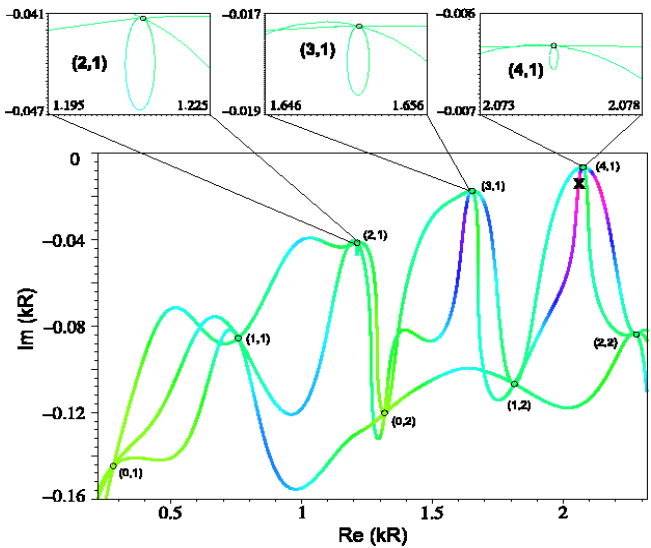}
}
\caption{(color online). Level dynamics of the resonances in the complex wavenumber plane for
a dielectric disk with $n=3.0$ and $R=1$ $\mu\text{m}$  and a point
scatterer of varying coupling parameter $a$.  
The quantum numbers of the unperturbed resonances are marked in brackets.
For the upper curve the scatterer has distance $d=0.99$ $\mu\text{m}$ from the
centre, 
for the middle curve $d=0.495$ $\mu$m, and for the lower curve $0.25$ $\mu$m. 
The color code indicates the  directionality of the emission 
(green marks small values of  $\Delta_{|f|^2}$, red marks high values of $\Delta_{|f|^2}$). 
The cross corresponds to the HD mode ($\Delta_{|f|^2}=2.12$, $Q=251$) 
with $kR=2.0571-\ui\, 0.0164 $ ($d=0.495$ $\mu$m, $a \approx 0.754$).} 
\label{Fig2}
\end{figure}

\begin{figure}[floatfix]
\centerline{
\includegraphics[width=8.6cm,height=5.3cm]{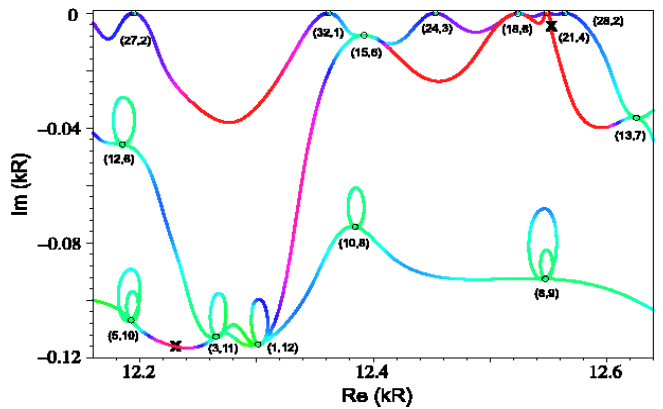}
}
\caption{ (color online). Continuation of Fig.~\ref{Fig2} for a different frequency range.
The lower cross corresponds to the HD mode ($\Delta_{|f|^2}=2.36$, $Q=212$) 
with $kR=12.2257-\ui\,0.1156$ ($d=0.25$ $\mu$m, $a \approx 0.047$). 
The upper cross corresponds to the HD mode ($\Delta_{|f|^2}=4.71$, $Q=15700$) 
with $kR=12.5513-\ui\,0.0016$ ($d=0.495$ $\mu$m, $a \approx 0.002$).}
\label{Fig3}
\end{figure}

We study the dynamics of the resonances upon varying the coupling parameter
$a$ for a typical GaAs microdisk
of effective refractive index $n=3$ and radius $R=1$ $\mu$m \cite{Peter2005},
with a point scatterer placed at three different distances
(0.99 $\mu$m, 0.495 $\mu$m, 0.25 $\mu$m) from the center of the disk. Figures~\ref{Fig2}~and~\ref{Fig3}
show the corresponding resonances in 
the frequency regions
$\nu 
 = {{c\,{\mathop{\rm Re}\nolimits} \left( k \right)} \mathord{\left/
 {\vphantom {{c{\mathop{\rm Re}\nolimits} \left( k \right)} {2\pi }}} \right.
 \kern-\nulldelimiterspace} {2\pi }}
=  0.048 - 1.098\times 10^{14} $ Hz 
(mid and near infrared) 
and $\nu = 5.806-6.035 \times 10^{14}$ Hz (green light), respectively.
Interestingly, the line segments parametrized by the coupling parameter $a$ do not only connect
different resonances of the unperturbed microdisk but as indicated in
the insets in Fig.~\ref{Fig2} and in Fig.~\ref{Fig3} there are also loops connecting single
resonances to themselves.

In order to quantify the far-field directionality of the electric field we consider
its asymptotic behaviour for $r\rightarrow\infty$ which has the form
\begin{equation}
E_z\left( {\bf r}, k_{{\rm res}} \right) = E_z\left(r,\varphi, k_{{\rm res}} \right) 
\propto \frac{{\exp (ik_{{\rm res}} r)}}{{\sqrt r }}f\left( \varphi  \right). \nonumber
\end{equation}
To characterize the directionality we compute the normalized
variance of the far-field intensity
\begin{equation}
\Delta_{|f|^2} = 
\int\limits_0^{2\pi } {| {f( \varphi )}|^4 d\varphi }  
/
\big( {\int\limits_0^{2\pi } {\left| {f( \varphi)} \right|^2 d\varphi } } \big)^2 -1\,.\nonumber
\end{equation}
From this definition it follows that
$\Delta_{|f|^2}=0$ and $\Delta_{|f|^2}=0.5$  for resonances which have $m=0$
and $m\ne0$, respectively, for $a=0$ or $a=\infty$.

In Figs.~\ref{Fig2}~and~\ref{Fig3} the directionality $\Delta_{|f|^2}$ is indicated by color
in the range from light green (low directional modes) 
through blue and violet to deep red (highly directional modes). 
It reaches values as high as $\Delta_{|f|^2} \approx 5$ for some specific modes.
We see that there are highly directional modes (HD) for a wide range of $Q$-factors
which in terms of the complex wavenumbers $k$ are given by 
$Q=-{\rm Re}(k)/{(2\,\rm Im}(k))$.

\begin{figure}[floatfix]
\centerline{
\includegraphics[width=7.6cm]{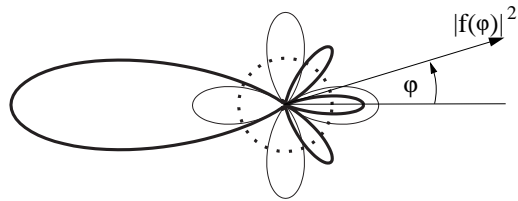}
}
\caption{Polar plot of the far-field intensity  $\left| {f\left( \varphi  \right)} \right|^2$  for
the unperturbed resonant mode $(0,1)$ (dashed circle),
the unperturbed resonant mode $(2,2)$ (thin solid line),
and the HD perturbed resonant mode marked by the cross in Fig.~\ref{Fig2} (thick solid line).
The parameters are the same as in Fig.~\ref{Fig2}.} 
\label{Fig4}
\end{figure}

\begin{figure}[floatfix]
\centerline{
\includegraphics[width=8.3cm]{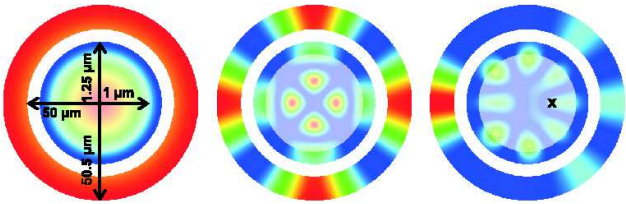}
}
\caption{ (color online). The intensity of the electric field in near- and far-field regions
for unperturbed resonant modes  $(0,1)$ (left), $(2,2)$ (middle),
and for the HD perturbed resonant mode marked by the cross in Fig.~\ref{Fig2} (right). 
The parameters are the same as in Fig.~\ref{Fig2}.} 
\label{Fig5}
\end{figure}

To illustrate the directionality in more detail we compare in 
Fig.~\ref{Fig4} the function ${\left| {f\left( \varphi  \right)} \right|^2}$ 
for two unperturbed resonant modes and
one perturbed HD mode in the frequency region of Fig.~\ref{Fig2}.
The corresponding near- and far-field electric field intensities 
are shown in Fig.~\ref{Fig5}. For a vanishing coupling strength
the angular dependence is sinusoidally modulated with the number of
minima being given by the quantum number $m$, 
i.e. only for $m=0$ is the emission isotropic. 
The radial quantum number $q$ determines 
the number of minima in the radial direction.

\begin{figure}[floatfix]
\centerline{       
\includegraphics[width=7.6cm]{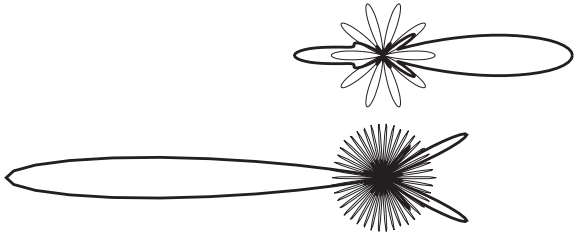}
}
\caption{  Polar plot of the far-field intensity  ${\left| {f\left( \varphi
        \right)} \right|^2}$ for 
the unperturbed resonance with quantum numbers $(m,q)=(5,10)$ (thin solid line
in the top panel),
the HD perturbed resonant mode marked by the lower cross in Fig.~\ref{Fig3}  (thick solid
line in the top panel),
the unperturbed resonance with   $(m,q)=(21,4)$ (thin solid line in the bottom
panel), the 
HD perturbed resonant mode marked by the upper cross in Fig.~\ref{Fig3} (thick solid
line in the bottom panel).  
The parameters are the same as in Fig.~\ref{Fig3}.} 
\label{Fig6}
\end{figure}

\begin{figure}[floatfix]
\centerline{
\includegraphics[width=5.5cm]{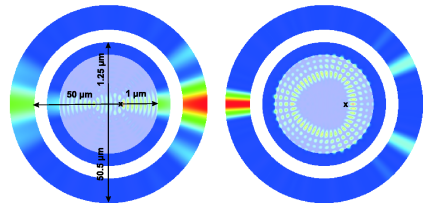}
}
\caption{(color online). The intensity of the electric field in near- and far-field regions for 
HD perturbed resonant mode shown by the lower cross in Fig.~\ref{Fig3} (left) 
and for the
HD perturbed resonant mode marked by the upper cross in Fig.~\ref{Fig3} (right). 
The  parameters are the same as  in Fig.~\ref{Fig3}.} 
\label{Fig7}
\end{figure}

Figures~\ref{Fig6}~and \ref{Fig7} show  the near- and far-field electric field intensities 
for two perturbed HD modes in the higher frequency region of Fig.~\ref{Fig3}.
This demonstrates that these can differ significantly from the nearby modes of
the unperturbed system also shown in Fig.~\ref{Fig6}.
While the near-field pattern of the mode shown in the right panel of
Fig.~\ref{Fig7} is still reminiscent of a WGM, the mode in the left panel is
strongly scarred. 
The far-field patterns shown in Figs.~\ref{Fig6}~and~\ref{Fig7} also show that
the scatterer can cause directional output in either direction of the symmetry axis.
The mode in the bottom panel of Fig.~\ref{Fig6} has both extremely high directionality and a very
high $Q$-factor. From Fig.~\ref{Fig3} we see that there is quite
a broad spectral range of green light with these properties the realization of 
which is a major goal in semiconductor physics.


{\em Conclusions.---}
In summary, we presented theoretical results that
demonstrate the existence of highly directional TM-modes
in the emission spectrum of a microdisk cavity
with a point scatterer. These modes
can appear even for a scatterer with a very weak
coupling constant which promises the feasibility of an experimental
realization of such cavities.  The level dynamics of the resonances upon
varying the coupling constant is of
theoretical interest in its own right and deserves further investigation.
Similarly, it would be interesting to get a deeper insight into the
output directionality by relating the resonance wavefunctions to the
underlying ray dynamics in the semiclassical limit.


\end{document}